%% file: main.tex
\documentclass[twocolumn,10pt, superscriptaddress, amsmath,amssymb, aps, prl, floatfix]{revtex4-2}

\usepackage{natbib}
\usepackage[normalem]{ulem}
\usepackage{comment}
\usepackage{graphicx}
\usepackage{dcolumn}
\usepackage{bm}
\usepackage{physics}
\usepackage{mathtools}
\usepackage{tensor}
\usepackage{xcolor}
\usepackage{booktabs}
\usepackage{multirow}
\usepackage{hyperref}
\usepackage[mathlines]{lineno}
\usepackage{amsmath}
\usepackage{enumitem}
\usepackage{color}
\usepackage{float}

\newcommand{\priorname}{\textsc{Goldilocks Prior}}
\newcommand{\lt}{\tilde{\Lambda}}
\newcommand{\MTOV}{m^{\text{max}}_{\text{TOV}}}
\newcommand{\dlt}{\delta\tilde{\Lambda}}
\newcommand{\mc}{\mathcal{M}_c}
\newcommand{\msol}{\text{M}_{\odot}}

\begin{document}

\title{Reanalysis of GW170817 and GW190425 under the Goldilocks Prior}

\author{Sanika Khadkikar}
\email[]{sanika@psu.edu}
\affiliation{Institute for Gravitation and the Cosmos, The Pennsylvania State University, University Park, PA 16802, USA}
\affiliation{Department of Physics, The Pennsylvania State University, University Park, PA 16802, USA}

\author{B. S. Sathyaprakash}
\affiliation{Institute for Gravitation and the Cosmos, The Pennsylvania State University, University Park, PA 16802, USA}
\affiliation{Department of Physics, The Pennsylvania State University, University Park, PA 16802, USA}
\affiliation{Department of Astronomy and Astrophysics, The Pennsylvania State University, University Park, PA 16802, USA}

\begin{abstract}
Gravitational waves from binary neutron star mergers encode the properties of the densest matter in the universe, but recovering them accurately requires controlling biases introduced by the inference framework. The recently developed \priorname\ \cite{khadkikar_goldilocks_prior} incorporates correlations predicted by a large ensemble of candidate equations of state and broadens them according to the expected detector sensitivity. We apply this framework to GW170817 and GW190425 and find that the inferred component tidal deformabilities differ substantially from standard analyses. For GW170817, we infer both neutron stars to be more tidally deformable than previously reported, with the most probable value for the lighter star nearly doubling from $\Lambda_2\simeq215$ to $\Lambda_2\simeq443$ relative to the uniform-prior analysis. For GW190425, the \priorname\ considerably reduces the range of allowed tidal deformabilities. We propagate the resulting event likelihoods to the dense matter equation of state using two complementary hierarchical methods, a strongly parametrized spectral representation and a broader semi-parametric Gaussian-process ensemble. For GW170817, both prior treatments yield consistent EoS constraints despite substantial differences in the inferred properties of the individual neutron stars. For GW190425, however, the EoS constraints differ beyond numerical scatter even after removing the source-level priors, showing that prior choice affects likelihood recovery for low-SNR events. The joint EoS constraints nevertheless remain largely unchanged because GW170817 dominates the inference.
\end{abstract}

\maketitle
\emph{\textbf{Introduction.~}}Neutron stars (NSs) contain the densest matter in the universe \cite{Lattimer:2012nd,Lattimer:2015nhk,Oertel:2016bki,Paschalidis:2016vmz,Ozel:2016oaf,Baym:2017whm}. Since NSs were first proposed in 1934, understanding their internal structure has been a central problem in nuclear physics and astrophysics \cite{Baade:1934zex}. Multimessenger observations of NSs over the last several decades have provided increasingly informative constraints on the distributions of their bulk properties, including the mass $m$, radius $R$, and the tidal deformability $\Lambda$, which quantifies the extent to which a NS deforms in an external tidal field \cite{Prakash:1996xs,Lattimer:2015nhk,Flanagan:2007ix,Hinderer:2007mb}. These properties can be inferred from the same observation or by combining information from different observational channels, but they are not independent. They are linked by the dense matter equation of state (EoS), which connects the microscopic physics of matter inside a NS to its macroscopic structure \cite{Prakash:1996xs,Lattimer:2015nhk}. Any inference of NS properties from observations must therefore account for these physical correlations.

Gravitational waves (GWs) from binary neutron star (BNS) mergers encode information about the component masses $m_i$ and tidal deformabilities $\Lambda_i$, where $i=1,2$ labels the two NSs in the binary, providing a way to constrain the EoS \cite{Schutz:1999xj,Read:2009yp,Farr:2015lna,Chatziioannou:2020pqz}. The LIGO-Virgo-KAGRA (LVK) GW detector network \cite{LIGOScientific:2014pky,VIRGO:2014yos,KAGRA:2020tym} has reported two events consistent with the BNS hypothesis: GW170817 \cite{LIGOScientific:2018mvr} and GW190425 \cite{LIGOScientific:2020aai}. Extracting the distributions of $m_i$ and $\Lambda_i$ characterizing these signals requires a statistical inference framework. The assumptions built into that framework can influence the inference of the source parameters and their statistical uncertainties. This is especially important for $\Lambda_i$, since tidal effects enter the GW signal at higher post-Newtonian order and are therefore more weakly constrained than the masses \cite{Flanagan:2007ix,Hinderer:2007mb,Favata:2013rwa}. Biases introduced at the inference stage can consequently propagate into the NS properties and EoS constraints derived from such GW observations.

The inference of source parameters from a GW signal requires a waveform model and prior distributions on its parameters. These priors describe the range of parameter values considered plausible before examining the data \cite{Finn:1992wt}. Since the NS EoS remains uncertain, standard GW analyses commonly adopt broad uniform priors on the tidal parameters to avoid favouring any particular EoS \cite{LIGOScientific:2018mvr, LIGOScientific:2020aai}. However, uniform priors in one set of parameters do not necessarily remain uniform in another. The leading and next-to-leading tidal contributions to the waveform model are linear, mass-weighted combinations of $\Lambda_i$'s termed as $\tilde{\Lambda}$ and $\delta\tilde{\Lambda}$ \cite{Blanchet:2013haa, Flanagan:2007ix, Hinderer:2007mb}. Uniform priors on $\Lambda_1$ and $\Lambda_2$ do not generally produce uniform distributions in these combinations. Similarly, uniform priors on $\tilde{\Lambda}$ and $\delta\tilde{\Lambda}$ acquire nonuniform effective distributions when we impose $\Lambda_i\geq0$ \cite{khadkikar_goldilocks_prior}. 

One can change the assumed prior after inference by reweighting the posterior samples. Let $\theta$ denote the set of waveform parameters and $d$ the observed GW data. Given a posterior $p_1(\theta\mid d)$ obtained under a prior $\pi_1(\theta)$, the posterior corresponding to a different prior $\pi_2(\theta)$ is
\begin{equation}
p_2(\theta\mid d) \propto p_1(\theta\mid d)
\frac{\pi_2(\theta)}{\pi_1(\theta)}.
\end{equation}
However, reweighting cannot recover regions where the original prior has no support. Even when the original prior has support, too few samples in a region can make the reweighted posterior unreliable \cite{khadkikar_goldilocks_prior, Prasad:2026idz}. Uniform priors on either $(\Lambda_1,\Lambda_2)$ or $(\tilde{\Lambda},\delta\tilde{\Lambda})$ can exclude physically plausible regions or assign them very low probabilities. As a result, these regions may be missed or poorly sampled during inference.

Beyond these limitations, uniform priors also ignore the physical relations between NS properties. The tidal deformability of a NS depends strongly on its mass and radius, with $\Lambda=\frac{2}{3}k_2(c^2R/Gm)^5$, where $c$ is the speed of light, $G$ is the gravitational constant, and the tidal Love number $k_2$ depends on the EoS \cite{Hinderer:2007mb}. For a given EoS, this establishes a relation between $m$ and $\Lambda$ that standard analyses do not generally account for. Recent studies have made considerable efforts to incorporate EoS information into tidal priors. \citet{Magnall:2025zhm} constructed an EoS-informed prior on $(\mc,q,\lt)$ but left out $\dlt$, which also carries information about the EoS. \citet{Wouters:2025csq} used normalizing flows to learn the joint distribution of masses and tidal deformabilities from parametrized EoSs. Even when all tidal parameters are included, parametrized EoS models can be overly restrictive. Their assumed functional forms can introduce artificial correlations between densities and exclude physically plausible EoSs \cite{Landry:2018prl}. Such priors can therefore bias the inference, particularly when the data are noisy.

We introduced the \priorname\ \cite{khadkikar_goldilocks_prior} to address these limitations together. 
It is a four-dimensional joint prior in $(\mc,q,\lt,\dlt)$, where $\mc$ is the chirp mass and $q=m_2/m_1\leq1$ is the mass ratio. We construct this prior using a broad semi-parametric ensemble of candidate EoSs. The ensemble combines nuclear physics constraints at low densities with flexible Gaussian-process extensions at higher densities, allowing a wider range of NS properties without assuming a single EoS functional form \cite{Ng:2025wdj}. We use this ensemble to capture the physical correlations between masses and tidal deformabilities, including those involving $\delta\tilde{\Lambda}$. We then broaden these correlations according to the expected measurement uncertainty of the detector network. This preserves support beyond the EoS ensemble and allows us to sample physically motivated regions directly rather than relying on reweighting. Tests using simulated BNS signals spanning different masses, tidal deformabilities, spins, and signal strengths showed consistently reduced biases relative to uniform and more restrictive EoS-informed priors \cite{khadkikar_goldilocks_prior}.

In this Letter, we reanalyze GW170817 and GW190425 using the \priorname, the standard implied uniform prior, and its reweighted counterpart. We first compare the inferred component tidal deformabilities and then recover the GW likelihoods for EoS inference. We carry out the latter with a spectral representation and a semi-parametric Gaussian-process ensemble \cite{Lindblom:2010bb, Ng:2025wdj}, which allow different ranges of EoS behavior. This provides two tests of the prior, one at the level of individual NS properties and another after the GW information has been mapped to the EoS.

\emph{\textbf{Source Parameters and Setup for EoS Inference.~}}We analyze the public strain data for GW170817 and GW190425 using the precessing tidal waveform model \textsc{IMRPhenomXP\_NRTidalv3}\cite{Pratten:2020fqn, Pratten:2020ceb, Abac:2023ujg}. For each event, we infer $m_i$, $\Lambda_i$, and the dimensionless spin magnitudes $\chi_i\equiv c|\vec{S}_i|/(Gm_i^2)$, where $\vec{S}_i$ is the spin angular momentum of the $i$th NS, along with the extrinsic binary parameters. Our fiducial analyses adopt low spin priors with $\chi_i\leq0.1$ following previous studies \cite{LIGOScientific:2018mvr}. For the standard analysis, we sample uniformly in $(\lt,\dlt)$, the tidal combinations entering the waveform phase at leading and next-to-leading orders. Imposing $\Lambda_i\geq0$ induces a nonuniform effective prior on $\lt$, allowing us to examine the effects of reweighting. Sampling uniformly in $(\Lambda_1,\Lambda_2)$ would not require reweighting to obtain uniform component priors and would give qualitatively similar results \cite{khadkikar_goldilocks_prior}. We therefore compare three treatments: (i) the standard uniform prior with the physicality constraint, (ii) its posterior reweighted to approximate a flat $\lt$ prior over the sampled region, and (iii) direct parameter estimation using the \priorname. We report the low-spin, precessing results here and provide the other spin and waveform configurations in the End Matter.

\begin{figure}[t]
    \centering
    \includegraphics[width=\linewidth]{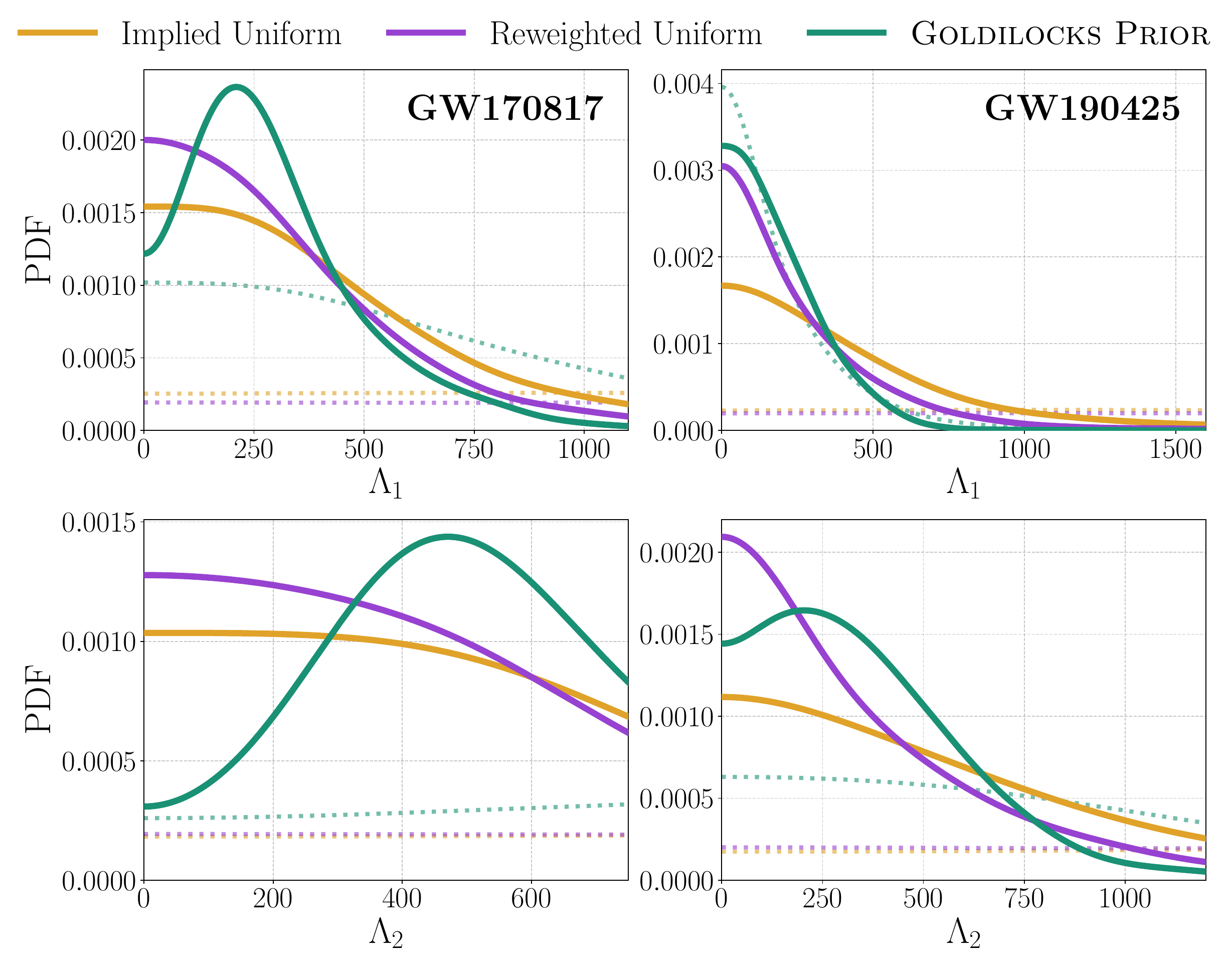}
    \caption{Posterior distributions of the component tidal deformabilities $\Lambda_1$ (top) and $\Lambda_2$ (bottom) for GW170817 (left) and GW190425 (right). Orange, purple, and teal curves correspond to the implied uniform prior, its reweighted counterpart, and the \priorname, respectively. Dotted curves show the corresponding prior distributions. For GW170817, the \priorname\ reduces the high-$\Lambda_1$ tail and shifts the $\Lambda_2$ peak toward larger values. 
    For GW190425, the \priorname\ narrows the $\Lambda_2$ posterior, while $\Lambda_1$ remains peaked near zero.}
    \label{fig:tidal}
\end{figure}

We then constrain the EoS using the likelihood distributions recovered for $m_i$ and $\Lambda_i$. This is carried out with the hierarchical inference framework \texttt{jester}, in which the per-event likelihoods for GW170817 and GW190425 are combined under the assumption that both systems are described by a common EoS \cite{Wouters:2025zju}. We perform this inference using two complementary EoS representations. The first uses a spectral parametrization, which restricts the EoS to a relatively compact functional family \cite{Lindblom:2010bb}. The second uses a semi-parametric ensemble that combines low-density nuclear constraints with Gaussian-process extensions at higher densities, allowing for a broader range of EoS behavior \cite{Ng:2025wdj}. 
We compare the two approaches to test whether the inferred shifts persist when the flexibility of the EoS representation is changed.

\begin{figure*}
    \centering
    \includegraphics[width=0.8\linewidth]{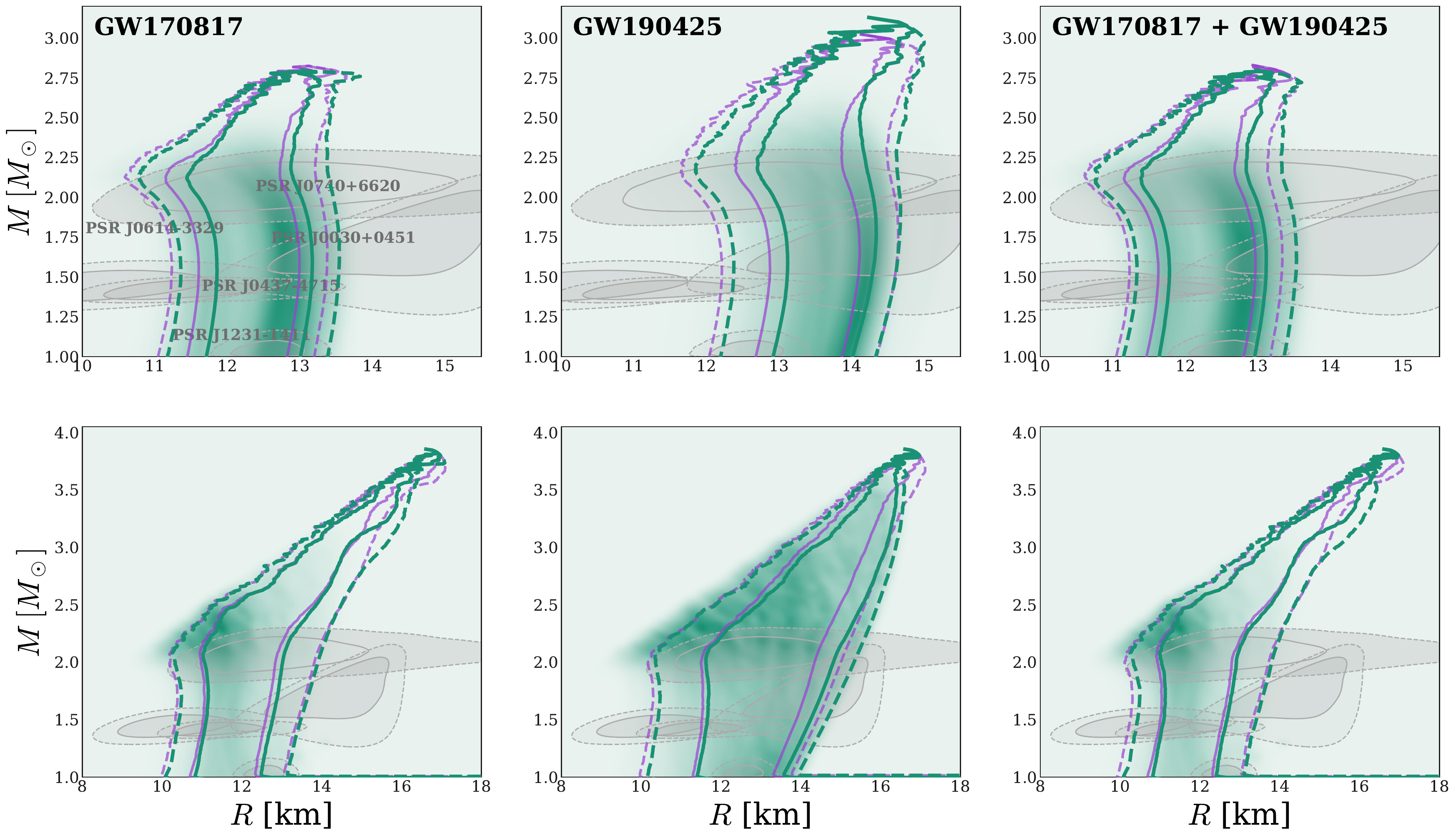}
    \caption{
    Mass--radius posteriors inferred from GW170817 (left), GW190425 (middle), and their joint analysis (right), using the likelihoods recovered from the uniform-prior (purple) and \priorname\ (teal) parameter-estimation runs. The top panels use the spectral EoS representation and the bottom panels the semi-parametric ensemble. Both priors yield consistent constraints for GW170817, while GW190425 shows a larger shift toward higher radii under the Goldilocks Prior. The joint constraints remain dominated by GW170817. Grey contours show NICER $m$--$R$ constraints for PSR J0030+0451 \cite{Vinciguerra:2023qxq}, PSR J0437--4715 \cite{Choudhury:2024xbk}, PSR J0614--3329 \cite{Mauviard:2025dmd}, PSR J1231--1411 \cite{Salmi:2024bss}, and PSR J0740+6620 \cite{Dittmann:2024mbo}.
    }
    \label{fig:mr}
\end{figure*}

For the hierarchical EoS analysis, we recover the GW likelihood from each event-level posterior by reweighting out the corresponding source-level prior. We refer to the likelihoods recovered from the \priorname\ and uniform-prior analyses as ``Reweighted Goldilocks'' and ``Reweighted Uniform,'' respectively. The hierarchical inference therefore uses the recovered GW likelihoods rather than the source-level posteriors. In both EoS analyses, we also include a likelihood constraint from the radio-timing mass measurement of PSR J0740+6620, $m=(2.08\pm0.07)\,\msol$ \cite{Fonseca:2021wxt}. This disfavors EoSs that cannot support a neutron star of the measured mass. Details of the \priorname\ construction and validation are given in the companion paper.

\emph{\textbf{Inference from Gravitational Wave Events.~}}Figure~\ref{fig:tidal} compares the component tidal deformabilities under the implied uniform prior, its reweighted counterpart, and the \priorname. For GW170817, the largest change is in $\Lambda_2$. Its posterior $p(\Lambda_2)$ peaks at $\Lambda_2\simeq215$ under the implied uniform prior and at $\Lambda_2\simeq443$ under the \priorname. The $90\%$ highest-posterior-density (HPD) interval changes from $\Lambda_2\in[2,1127]$ to $\Lambda_2\in[111,1150]$. The implied uniform posterior on $\Lambda_1$ is nearly flat for $\Lambda_1\lesssim200$, whereas the \priorname\ produces a peak near $\Lambda_1\simeq190$ and reduces the upper HPD bound from $\Lambda_1=855$ to $\Lambda_1=606$. For two NSs following the same EoS, $m_1\geq m_2$ implies $\Lambda_1\leq\Lambda_2$. This relation is built into the \priorname\ but not the independent tidal priors used in the uniform analysis. The GW170817 posteriors remain distinct from the corresponding priors (dotted curves in Fig.~\ref{fig:tidal}), showing that the inferred distributions are significantly informed by the data itself.

GW190425 provides a complementary test case. Its chirp mass, $\mc\simeq1.44\,\msol$, is substantially higher than that of GW170817 ($\mc\simeq1.19\,\msol$), so smaller component tidal deformabilities are expected for a given EoS. Under the \priorname, the $\Lambda_1$ posterior peak shifts from $\simeq132$ to $\simeq59$, while its $90\%$ HPD upper bound decreases from 1024 to 397. The upper bound on $\Lambda_2$ also decreases substantially, from 1478 to 698. Both posteriors retain support at zero. These reductions are consistent with the expected decrease in tidal deformability with increasing NS mass. Unlike the uniform prior, the \priorname\ incorporates this EoS-predicted correlation between mass and tidal deformability.

\emph{\textbf{Equation of State Constraints.~}}We infer the EoS using the ``Reweighted Uniform'' and ``Reweighted Goldilocks'' likelihood distributions under the spectral and semi-parametric EoS representations. Figure~\ref{fig:mr} shows the resulting $m$--$R$ posteriors. For GW170817, the higher SNR ensures that both source-level analyses adequately sample the regions relevant to EoS inference. Reweighting therefore recovers consistent likelihood distributions, giving similar EoS constraints under both representations despite differences in the source-level $\Lambda_i$ posteriors.

For GW190425, the lower SNR produces a broad likelihood distribution that admits combinations of $(m_i,\Lambda_i)$ that fit the GW data but are inconsistent with physically plausible NS EoSs. The uniform prior admits these configurations, whereas the \priorname\ suppresses them through the EoS-predicted $m$--$\Lambda$ correlations. Although reweighting removes the source-level priors, it cannot recover regions inadequately sampled during event-level inference, leading to differences in the reconstructed likelihood distributions. The \priorname\ reconstruction consequently raises the lower $R$ bound in Fig.~\ref{fig:mr}. The effect also depends on the EoS representation. The spectral parametrization introduces correlations between $P(\rho)$ at different densities, restricting the range of EoSs it can represent. The semi-parametric ensemble retains greater freedom at high densities, including more configurations with large $\Lambda_i$ at higher $m_i$ \cite{Lindblom:2010bb,Ng:2025wdj}. Differences between the reconstructed likelihood distributions are therefore more apparent at larger $R$ in the semi-parametric $m$--$R$ posteriors.

\begin{figure*}[t]
    \centering
    \includegraphics[width=\linewidth]{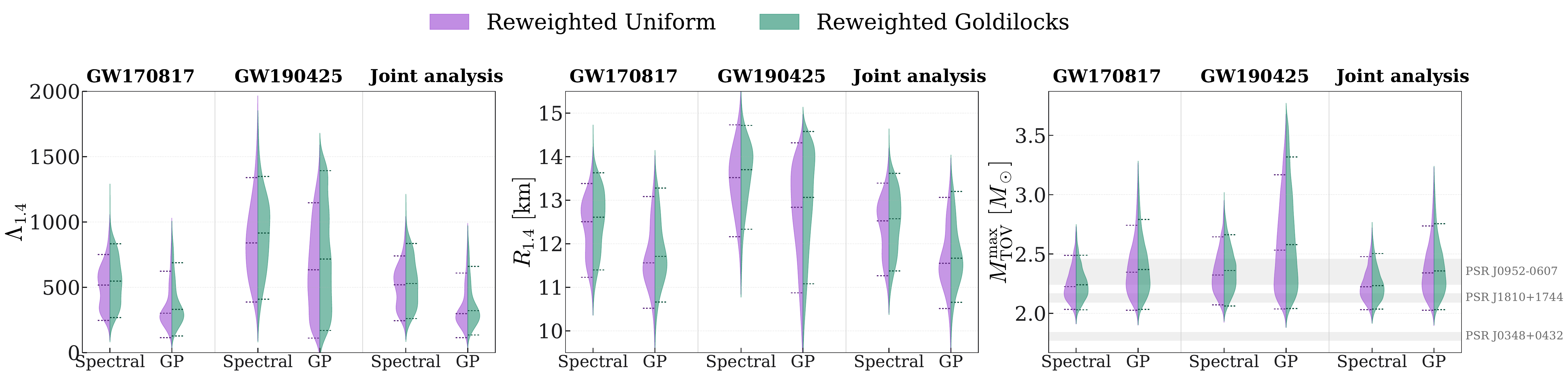}
    \caption{
    Posterior distributions of $\Lambda_{1.4}$ (left), $R_{1.4}$ (middle), and $m_{\rm TOV}^{\max}$ (right) for GW170817, GW190425 and the joint analysis of both together. Purple and teal violins show the hierarchical EoS posteriors obtained using the likelihoods recovered from the uniform and \priorname\ analyses, respectively. These are labeled ``Reweighted Uniform'' and ``Reweighted Goldilocks'' because the corresponding source-level priors have been removed before EoS inference. Each grouping shows results from the spectral and semi-parametric Gaussian-process EoS representations. Horizontal markers indicate the medians and $2\sigma$ credible intervals. Grey bands in the right panel show the measured masses of PSR J0952--0607, $m=(2.35\pm0.11)\,\msol$ \cite{Romani:2025ytn}, PSR J1810+1744, $m=(2.13\pm0.04)\,\msol$ \cite{Romani:2021xmb}, and PSR J0348+0432, $m=(1.806\pm0.037)\,\msol$ \cite{Saffer:2024tlb}.
    }
    \label{fig:violin}
\end{figure*}

\emph{\textbf{Canonical NS Properties.~}}Figure~\ref{fig:violin} shows the posteriors on $\Lambda_{1.4}$, $R_{1.4}$, and $\MTOV$ for GW170817, GW190425, and their joint analysis. For GW170817, both likelihood reconstructions give closely overlapping posteriors. The median $R_{1.4}$ increases from $12.50$ km to $12.61$ km in the spectral representation and from $11.56$ km to $11.71$ km in the semi-parametric representation. The posterior distributions remain largely unchanged, consistent with the agreement between the reconstructed likelihood distributions discussed above.

For GW190425, the changes are more pronounced and extend beyond the medians. Under Reweighted Goldilocks, the $R_{1.4}$ posterior shifts toward larger values in both EoS representations. In the spectral representation, the median $R_{1.4}$ increases from $13.52$ km to $13.70$ km, while its $90\%$ HPD interval changes from $[12.16,14.73]\,\mathrm{km}$ to $[12.33,14.72]\,\mathrm{km}$. In the semi-parametric representation, the median $R_{1.4}$ increases from $12.83$ to $13.07$ km, with its $90\%$ HPD interval shifting from $[10.87,14.31]\,\mathrm{km}$ to $[11.09,14.58]\,\mathrm{km}$. The median $\Lambda_{1.4}$ also increases from 839 to 916 in the spectral representation and from 634 to 717 in the semi-parametric representation. These changes extend across the posterior distributions rather than being limited to their central estimates.

Adding GW190425 to GW170817 nevertheless changes the joint constraints very little. The joint median $\Lambda_{1.4}$ changes from 519 to 529 in the spectral representation, while the $R_{1.4}$ and $\MTOV$ posteriors remain consistent. However, GW190425 illustrates an important limitation of EoS inference from low-SNR events. Although the likelihood function is independent of the source-level prior, the inferred likelihood distribution from finite posterior samples can retain sensitivity to that prior even after reweighting. The \priorname\ improves sampling of physically plausible $(m_i,\Lambda_i)$ configurations. As future GW catalogs include many such low SNR signals, accounting for these sampling effects will become increasingly important for reliable EoS inference.

\emph{\textbf{Conclusions.~}}We have reanalyzed GW170817 and GW190425 using the \priorname\ \cite{khadkikar_goldilocks_prior} and compared the inferred $(m_i,\Lambda_i)$ with those obtained under the uniform prior. We find substantial changes in the $\Lambda_i$ posteriors for both events. For GW170817, the allowed range of $\Lambda_1$ decreases, while the $\Lambda_2$ posterior favors larger values. For GW190425, both $\Lambda_i$ have considerably smaller upper bounds, consistent with its higher $\mc$ and the expected $m$--$\Lambda$ relation for a given EoS. These results show that the choice of prior can substantially change the inferred properties of individual NSs, even when analyzing the same GW data.

We then used the likelihood distributions recovered from both analyses to infer the EoS under the spectral and semi-parametric representations. For GW170817, reweighting out the source-level priors gives consistent EoS constraints despite the differences in $\Lambda_i$. Its higher SNR allows both analyses to adequately sample the regions relevant to EoS inference. For GW190425, the lower SNR makes the reconstructed likelihood distributions more sensitive to the regions sampled under each prior. The Reweighted Goldilocks analysis shifts the $R_{1.4}$ and $\Lambda_{1.4}$ posteriors toward larger values and gives a modest increase in $\MTOV$ under both EoS representations. Despite these differences, the joint constraints remain dominated by GW170817 and are largely unchanged by the addition of GW190425.

Future GW catalogs will contain many events with limited information about $\Lambda_i$. Combining these detections to constrain the EoS will require more than simply increasing the number of observations. Each event must contribute a reliably reconstructed likelihood distribution, particularly when the data cannot strongly constrain the individual $\Lambda_i$. The \priorname\ addresses this challenge by incorporating the EoS-predicted $m$--$\Lambda$ correlations while broadening its support according to detector sensitivity. Our results demonstrate the importance of this balance for recovering physically plausible NS properties and obtaining reliable EoS constraints from increasingly large GW catalogs.


\emph{\textbf{Acknowledgements.~}}For insightful discussions and feedback we thank K.~G.~Arun, Nils Andersson, Jolien Creighton, Nathan Johnson-McDaniel, and Jocelyn Read. We also thank Colm Talbot for help in the development of the joint prior templates implemented in \textsc{bilby}. This research is supported by National Science Foundation grants AST-2307147, PHY-2308886, PHY-2309064, PHY-2608011 and AST-2606673.


\bibliography{bibliography}

\clearpage
\onecolumngrid

\section{End Matter}
\footnotesize
\setlength{\tabcolsep}{4pt}

\emph{\textbf{Analysis Configurations.~}}Each event is analyzed with \textsc{bilby} and the \textsc{dynesty} sampler under four configurations: two
waveform models, IMRPhenomXP\_NRTidalv3 (precessing, ``XP'') and IMRPhenomXAS\_NRTidalv3 (aligned-spin,
``XAS''), and two spin priors, $\chi_i \le 0.1$ (``low'') and $\chi_i \le 0.89$ (``high''). The main text
quotes the low-spin XP configuration. Table~\ref{tab:config} lists the settings common to the uniform and
Goldilocks analyses of each event. The uniform analysis samples $\tilde\Lambda$ and $\delta\tilde\Lambda$
uniformly with the constraint $0 \le \Lambda_i \le 5000$; the Goldilocks analysis replaces the priors on
$(\mathcal{M}_c, q, \tilde\Lambda, \delta\tilde\Lambda)$ by the four-dimensional joint prior of \citet{khadkikar_goldilocks_prior} with
mirrored boundary handling. All other priors are shared.

\begin{table}[H]
\centering
\caption{Parameter estimation settings. $f_{\min}$ is the lower frequency cutoff, $f_{\rm ref}$ the reference
frequency, and $n_{\rm live}$ the number of live points.}
\label{tab:config}
\begin{tabular}{lcc}
\toprule
 & GW170817 & GW190425 \\
\midrule
Segment duration [s] & 128 & 128 \\
$f_{\min}$ [Hz] & 23 & 20 \\
$f_{\max}$ [Hz] & 2048 & 2048 \\
$f_{\rm ref}$ [Hz] & 50 & 50 \\
Detector-frame $\mathcal{M}_c$ prior [$M_\odot$] & $[1.1838,\,1.22]$ & $[1.485,\,1.49]$ \\
$q$ prior & $[0.125,\,1]$ & $[0.125,\,1]$ \\
$\tilde\Lambda$ prior (uniform analysis) & $[0,\,5000]$ & $[0,\,5000]$ \\
$\delta\tilde\Lambda$ prior (uniform analysis) & $[-1000,\,1000]$ & $[-1000,\,1000]$ \\
$d_L$ prior [Mpc] & $[1,\,75]$ & $[10,\,1000]$ \\
$n_{\rm live}$ & 2048 & 2048 \\
Distance marginalization & yes & yes \\
\bottomrule
\end{tabular}
\end{table}

The hierarchical EoS inference uses \textsc{jester} with 4000 sequential Monte Carlo particles for the spectral
representation and importance weights over an ensemble of $10^5$ semi-parametric Gaussian-process EoSs. Both
include the mass measurement of PSR~J0740+6620. Event likelihoods are recovered from the posterior samples by
dividing out the full sampling prior, including the mass and tidal Jacobians, and are represented by a normalizing
flow trained on $3\times10^4$ resampled draws. Quoted medians and intervals for the spectral representation use
the final particle weights returned by the sampler.

\emph{\textbf{GW170817 Tidal Posteriors.~}}Table~\ref{tab:gw170817} summarizes the component tidal deformabilities for every configuration. The
posteriors are skewed with long tails, so we report the location of the posterior peak together with the 90\%
highest-posterior-density (HPD) interval rather than the median.

\begin{table}[H]
\centering
\caption{GW170817 tidal posteriors. Peaks are the maxima of a Gaussian kernel density estimate; intervals are 90\% HPD.}
\label{tab:gw170817}
\input{supplement_tables_gw170817}
\end{table}

\newpage
\emph{\textbf{GW190425 Tidal Posteriors.~}}
\begin{table}[H]
\centering
\caption{GW190425 tidal posteriors, as in Table~\ref{tab:gw170817}.}
\label{tab:gw190425}
\input{supplement_tables_gw190425}
\end{table}

\emph{\textbf{Additional Equation of State Results.~}}
Tables~\ref{tab:eos_spectral} and \ref{tab:eos_gp} give $\Lambda_{1.4}$, $R_{1.4}$, and $\MTOV$ for every combination of events, spin prior, waveform, and source-level prior, for the spectral and semi-parametric representations respectively. Values are medians with 90\% HPD intervals. The Uniform and Goldilocks entries refer to EoS inference using likelihoods recovered from the corresponding parameter-estimation runs after removing their sampling priors.

\begin{table}[H]
\begin{minipage}[t]{0.5\linewidth}
\caption{Spectral EoS representation.}
\label{tab:eos_spectral}
\centering\scriptsize\setlength{\tabcolsep}{1.2pt}
\input{supplement_tables_spectral}
\end{minipage}%
\begin{minipage}[t]{0.5\linewidth}
\caption{Semi-parametric Gaussian-process EoS representation.}
\label{tab:eos_gp}
\centering\scriptsize\setlength{\tabcolsep}{1.2pt}
\input{supplement_tables_gp}
\end{minipage}
\end{table}

\end{document}

%% file: supplement_tables_gw170817.tex
\begin{tabular}{lll rl rl c}
\toprule
Spin & Waveform & Prior & \multicolumn{2}{c}{$\Lambda_1$} & \multicolumn{2}{c}{$\Lambda_2$} & $q$ \\
 & & & peak & 90\% HPD & peak & 90\% HPD & median \\
\midrule
Low & XP & Uniform & 158 & $[0,\,855]$ & 215 & $[2,\,1127]$ & 0.87 \\
Low & XP & Goldilocks & 190 & $[27,\,606]$ & 443 & $[111,\,1150]$ & 0.87 \\
Low & XAS & Uniform & 182 & $[0,\,794]$ & 214 & $[0,\,1217]$ & 0.87 \\
Low & XAS & Goldilocks & 189 & $[1,\,569]$ & 443 & $[14,\,1057]$ & 0.87 \\
High & XP & Uniform & 101 & $[0,\,690]$ & 368 & $[0,\,1747]$ & 0.71 \\
High & XP & Goldilocks & 160 & $[1,\,505]$ & 458 & $[30,\,1428]$ & 0.83 \\
High & XAS & Uniform & 97 & $[0,\,687]$ & 257 & $[0,\,1448]$ & 0.77 \\
High & XAS & Goldilocks & 101 & $[0,\,493]$ & 467 & $[35,\,1473]$ & 0.79 \\
\bottomrule
\end{tabular}

%% file: supplement_tables_gw190425.tex
\begin{tabular}{lll rl rl c}
\toprule
Spin & Waveform & Prior & \multicolumn{2}{c}{$\Lambda_1$} & \multicolumn{2}{c}{$\Lambda_2$} & $q$ \\
 & & & peak & 90\% HPD & peak & 90\% HPD & median \\
\midrule
Low & XP & Uniform & 132 & $[0,\,1024]$ & 188 & $[0,\,1478]$ & 0.89 \\
Low & XP & Goldilocks & 59 & $[0,\,397]$ & 141 & $[0,\,698]$ & 0.89 \\
Low & XAS & Uniform & 133 & $[0,\,955]$ & 173 & $[0,\,1324]$ & 0.90 \\
Low & XAS & Goldilocks & 65 & $[0,\,394]$ & 184 & $[2,\,683]$ & 0.90 \\
High & XP & Uniform & 138 & $[0,\,998]$ & 247 & $[0,\,2058]$ & 0.77 \\
High & XP & Goldilocks & 41 & $[0,\,368]$ & 284 & $[0,\,1491]$ & 0.78 \\
High & XAS & Uniform & 158 & $[0,\,1317]$ & 304 & $[0,\,2326]$ & 0.78 \\
High & XAS & Goldilocks & 42 & $[0,\,383]$ & 284 & $[0,\,1416]$ & 0.79 \\
\bottomrule
\end{tabular}

%% file: supplement_tables_spectral.tex
\begin{tabular}{llll ccc}
\toprule
Events & Spin & Waveform & Prior & $\Lambda_{1.4}$ & $R_{1.4}$ [km] & $m^{\max}_{\rm TOV}$ [$M_\odot$] \\
\midrule
GW170817 & Low & XP & Uniform & $517^{+233}_{-271}$ & $12.50^{+0.88}_{-1.27}$ & $2.22^{+0.26}_{-0.19}$ \\
GW170817 & Low & XP & Goldilocks & $548^{+286}_{-279}$ & $12.61^{+1.02}_{-1.21}$ & $2.24^{+0.25}_{-0.21}$ \\
GW170817 & Low & XAS & Uniform & $543^{+250}_{-294}$ & $12.60^{+0.88}_{-1.34}$ & $2.23^{+0.26}_{-0.21}$ \\
GW170817 & Low & XAS & Goldilocks & $491^{+257}_{-247}$ & $12.40^{+1.02}_{-1.13}$ & $2.22^{+0.24}_{-0.20}$ \\
GW170817 & High & XP & Uniform & $526^{+236}_{-305}$ & $12.54^{+0.91}_{-1.44}$ & $2.23^{+0.26}_{-0.20}$ \\
GW170817 & High & XP & Goldilocks & $534^{+271}_{-285}$ & $12.57^{+1.00}_{-1.30}$ & $2.23^{+0.26}_{-0.21}$ \\
GW170817 & High & XAS & Uniform & $528^{+231}_{-292}$ & $12.54^{+0.88}_{-1.35}$ & $2.23^{+0.25}_{-0.21}$ \\
GW170817 & High & XAS & Goldilocks & $567^{+197}_{-314}$ & $12.70^{+0.68}_{-1.47}$ & $2.24^{+0.27}_{-0.20}$ \\
GW190425 & Low & XP & Uniform & $839^{+501}_{-452}$ & $13.52^{+1.21}_{-1.36}$ & $2.32^{+0.32}_{-0.25}$ \\
GW190425 & Low & XP & Goldilocks & $916^{+432}_{-507}$ & $13.70^{+1.02}_{-1.37}$ & $2.36^{+0.30}_{-0.30}$ \\
GW190425 & Low & XAS & Uniform & $835^{+474}_{-487}$ & $13.49^{+1.28}_{-1.29}$ & $2.33^{+0.32}_{-0.26}$ \\
GW190425 & Low & XAS & Goldilocks & $967^{+611}_{-516}$ & $13.81^{+1.26}_{-1.32}$ & $2.37^{+0.33}_{-0.29}$ \\
GW190425 & High & XP & Uniform & $881^{+545}_{-474}$ & $13.62^{+1.25}_{-1.28}$ & $2.35^{+0.30}_{-0.28}$ \\
GW190425 & High & XP & Goldilocks & $961^{+596}_{-545}$ & $13.81^{+1.27}_{-1.34}$ & $2.37^{+0.32}_{-0.29}$ \\
GW190425 & High & XAS & Uniform & $916^{+542}_{-501}$ & $13.70^{+1.23}_{-1.37}$ & $2.36^{+0.32}_{-0.28}$ \\
GW190425 & High & XAS & Goldilocks & $997^{+571}_{-525}$ & $13.89^{+1.16}_{-1.35}$ & $2.37^{+0.33}_{-0.28}$ \\
Combined & Low & XP & Uniform & $519^{+222}_{-275}$ & $12.52^{+0.87}_{-1.26}$ & $2.22^{+0.25}_{-0.19}$ \\
Combined & Low & XP & Goldilocks & $529^{+306}_{-268}$ & $12.57^{+1.04}_{-1.20}$ & $2.23^{+0.27}_{-0.20}$ \\
Combined & Low & XAS & Uniform & $546^{+234}_{-283}$ & $12.61^{+0.80}_{-1.33}$ & $2.24^{+0.25}_{-0.20}$ \\
Combined & Low & XAS & Goldilocks & $474^{+263}_{-246}$ & $12.33^{+1.10}_{-1.06}$ & $2.22^{+0.24}_{-0.19}$ \\
Combined & High & XP & Uniform & $543^{+222}_{-291}$ & $12.60^{+0.79}_{-1.38}$ & $2.25^{+0.25}_{-0.21}$ \\
Combined & High & XP & Goldilocks & $526^{+274}_{-269}$ & $12.53^{+1.01}_{-1.23}$ & $2.23^{+0.24}_{-0.21}$ \\
Combined & High & XAS & Uniform & $538^{+236}_{-291}$ & $12.58^{+0.88}_{-1.29}$ & $2.24^{+0.25}_{-0.21}$ \\
Combined & High & XAS & Goldilocks & $576^{+173}_{-330}$ & $12.72^{+0.70}_{-1.40}$ & $2.24^{+0.26}_{-0.19}$ \\
\bottomrule
\end{tabular}

%% file: supplement_tables_gp.tex
\begin{tabular}{llll ccc}
\toprule
Events & Spin & Waveform & Prior & $\Lambda_{1.4}$ & $R_{1.4}$ [km] & $m^{\max}_{\rm TOV}$ [$M_\odot$] \\
\midrule
GW170817 & Low & XP & Uniform & $302^{+321}_{-188}$ & $11.56^{+1.52}_{-1.04}$ & $2.35^{+0.40}_{-0.32}$ \\
GW170817 & Low & XP & Goldilocks & $331^{+357}_{-205}$ & $11.71^{+1.57}_{-1.05}$ & $2.37^{+0.42}_{-0.34}$ \\
GW170817 & Low & XAS & Uniform & $317^{+343}_{-176}$ & $11.65^{+1.45}_{-1.13}$ & $2.35^{+0.43}_{-0.33}$ \\
GW170817 & Low & XAS & Goldilocks & $315^{+280}_{-189}$ & $11.62^{+1.36}_{-1.10}$ & $2.35^{+0.37}_{-0.32}$ \\
GW170817 & High & XP & Uniform & $296^{+365}_{-176}$ & $11.52^{+1.58}_{-1.14}$ & $2.34^{+0.43}_{-0.32}$ \\
GW170817 & High & XP & Goldilocks & $303^{+319}_{-192}$ & $11.56^{+1.49}_{-1.17}$ & $2.34^{+0.40}_{-0.31}$ \\
GW170817 & High & XAS & Uniform & $289^{+323}_{-174}$ & $11.47^{+1.53}_{-1.10}$ & $2.33^{+0.39}_{-0.32}$ \\
GW170817 & High & XAS & Goldilocks & $309^{+354}_{-187}$ & $11.59^{+1.53}_{-1.16}$ & $2.34^{+0.41}_{-0.31}$ \\
GW190425 & Low & XP & Uniform & $634^{+514}_{-522}$ & $12.83^{+1.48}_{-1.96}$ & $2.53^{+0.63}_{-0.50}$ \\
GW190425 & Low & XP & Goldilocks & $717^{+675}_{-547}$ & $13.07^{+1.51}_{-1.98}$ & $2.58^{+0.74}_{-0.54}$ \\
GW190425 & Low & XAS & Uniform & $612^{+517}_{-498}$ & $12.78^{+1.55}_{-1.82}$ & $2.52^{+0.63}_{-0.48}$ \\
GW190425 & Low & XAS & Goldilocks & $759^{+695}_{-529}$ & $13.18^{+1.44}_{-1.96}$ & $2.61^{+0.77}_{-0.56}$ \\
GW190425 & High & XP & Uniform & $702^{+479}_{-577}$ & $13.02^{+1.39}_{-1.89}$ & $2.58^{+0.63}_{-0.53}$ \\
GW190425 & High & XP & Goldilocks & $766^{+700}_{-528}$ & $13.19^{+1.44}_{-1.92}$ & $2.63^{+0.74}_{-0.57}$ \\
GW190425 & High & XAS & Uniform & $758^{+515}_{-571}$ & $13.16^{+1.33}_{-1.75}$ & $2.63^{+0.63}_{-0.56}$ \\
GW190425 & High & XAS & Goldilocks & $902^{+631}_{-638}$ & $13.50^{+1.24}_{-2.08}$ & $2.71^{+0.80}_{-0.62}$ \\
Combined & Low & XP & Uniform & $299^{+310}_{-185}$ & $11.55^{+1.51}_{-1.04}$ & $2.34^{+0.40}_{-0.31}$ \\
Combined & Low & XP & Goldilocks & $321^{+339}_{-187}$ & $11.67^{+1.54}_{-1.01}$ & $2.36^{+0.40}_{-0.33}$ \\
Combined & Low & XAS & Uniform & $324^{+323}_{-184}$ & $11.67^{+1.37}_{-1.15}$ & $2.36^{+0.42}_{-0.33}$ \\
Combined & Low & XAS & Goldilocks & $312^{+276}_{-174}$ & $11.60^{+1.29}_{-1.08}$ & $2.35^{+0.35}_{-0.32}$ \\
Combined & High & XP & Uniform & $305^{+366}_{-186}$ & $11.58^{+1.55}_{-1.18}$ & $2.35^{+0.44}_{-0.33}$ \\
Combined & High & XP & Goldilocks & $308^{+298}_{-183}$ & $11.58^{+1.36}_{-1.20}$ & $2.35^{+0.39}_{-0.31}$ \\
Combined & High & XAS & Uniform & $319^{+341}_{-196}$ & $11.63^{+1.43}_{-1.25}$ & $2.37^{+0.43}_{-0.34}$ \\
Combined & High & XAS & Goldilocks & $308^{+351}_{-182}$ & $11.58^{+1.52}_{-1.14}$ & $2.34^{+0.41}_{-0.31}$ \\
\bottomrule
\end{tabular}

%% file: bibliography.bib
@unpublished{khadkikar_goldilocks_prior,
  author       = {Khadkikar, Sanika and Sathyaprakash, Bangalore},
  title        = {The Goldilocks Prior: Physically Consistent Inference of Neutron Star Properties from Gravitational Wave Observations},
  note         = {(in preparation)},
  institution  = {Institute for Gravitation and the Cosmos, Pennsylvania State University},
  year         = {2026}
}

@article{Prakash:1996xs,
    author = "Prakash, Madappa and Bombaci, Ignazio and Prakash, Manju and Ellis, Paul J. and Lattimer, James M. and Knorren, Roland",
    title = "{Composition and structure of protoneutron stars}",
    eprint = "nucl-th/9603042",
    archivePrefix = "arXiv",
    reportNumber = "SUNY-NTG-96-11, NUC-MINN-93-23-T",
    doi = "10.1016/S0370-1573(96)00023-3",
    journal = "Phys. Rept.",
    volume = "280",
    pages = "1--77",
    year = "1997"
}

@article{Lattimer:2012nd,
    author = "Lattimer, James M.",
    title = "{The nuclear equation of state and neutron star masses}",
    eprint = "1305.3510",
    archivePrefix = "arXiv",
    primaryClass = "nucl-th",
    doi = "10.1146/annurev-nucl-102711-095018",
    journal = "Ann. Rev. Nucl. Part. Sci.",
    volume = "62",
    pages = "485--515",
    year = "2012"
}

@article{Lattimer:2015nhk,
    author = "Lattimer, James M. and Prakash, Madappa",
    title = "{The Equation of State of Hot, Dense Matter and Neutron Stars}",
    eprint = "1512.07820",
    archivePrefix = "arXiv",
    primaryClass = "astro-ph.SR",
    doi = "10.1016/j.physrep.2015.12.005",
    journal = "Phys. Rept.",
    volume = "621",
    pages = "127--164",
    year = "2016"
}

@article{Ozel:2016oaf,
    author = {\"Ozel, Feryal and Freire, Paulo},
    title = "{Masses, Radii, and the Equation of State of Neutron Stars}",
    eprint = "1603.02698",
    archivePrefix = "arXiv",
    primaryClass = "astro-ph.HE",
    doi = "10.1146/annurev-astro-081915-023322",
    journal = "Ann. Rev. Astron. Astrophys.",
    volume = "54",
    pages = "401--440",
    year = "2016"
}

@article{Oertel:2016bki,
    author = {Oertel, M. and Hempel, M. and Kl\"ahn, T. and Typel, S.},
    title = "{Equations of state for supernovae and compact stars}",
    eprint = "1610.03361",
    archivePrefix = "arXiv",
    primaryClass = "astro-ph.HE",
    doi = "10.1103/RevModPhys.89.015007",
    journal = "Rev. Mod. Phys.",
    volume = "89",
    number = "1",
    pages = "015007",
    year = "2017"
}

@article{Paschalidis:2016vmz,
    author = "Paschalidis, Vasileios and Stergioulas, Nikolaos",
    title = "{Rotating Stars in Relativity}",
    eprint = "1612.03050",
    archivePrefix = "arXiv",
    primaryClass = "astro-ph.HE",
    doi = "10.1007/s41114-017-0008-x",
    journal = "Living Rev. Rel.",
    volume = "20",
    number = "1",
    pages = "7",
    year = "2017"
}

@article{Baym:2017whm,
    author = "Baym, Gordon and Hatsuda, Tetsuo and Kojo, Toru and Powell, Philip D. and Song, Yifan and Takatsuka, Tatsuyuki",
    title = "{From hadrons to quarks in neutron stars: a review}",
    eprint = "1707.04966",
    archivePrefix = "arXiv",
    primaryClass = "astro-ph.HE",
    reportNumber = "RIKEN-ITHEMS-REPORT-17, RIKEN-QHP-316, RIKEN-iTHEMS-Report-17",
    doi = "10.1088/1361-6633/aaae14",
    journal = "Rept. Prog. Phys.",
    volume = "81",
    number = "5",
    pages = "056902",
    year = "2018"
}

@article{Schutz:1999xj,
    author = "Schutz, Bernard F.",
    title = "{Gravitational wave astronomy}",
    eprint = "gr-qc/9911034",
    archivePrefix = "arXiv",
    reportNumber = "AEI-1999-34",
    doi = "10.1088/0264-9381/16/12A/307",
    journal = "Class. Quant. Grav.",
    volume = "16",
    pages = "A131--A156",
    year = "1999"
}

@article{Read:2009yp,
    author = "Read, Jocelyn S. and Markakis, Charalampos and Shibata, Masaru and Uryu, Koji and Creighton, Jolien D. E. and Friedman, John L.",
    title = "{Measuring the neutron star equation of state with gravitational wave observations}",
    eprint = "0901.3258",
    archivePrefix = "arXiv",
    primaryClass = "gr-qc",
    doi = "10.1103/PhysRevD.79.124033",
    journal = "Phys. Rev. D",
    volume = "79",
    pages = "124033",
    year = "2009"
}

@article{LIGOScientific:2014pky,
    author = "Aasi, J. and others",
    collaboration = "LIGO Scientific",
    title = "{Advanced LIGO}",
    eprint = "1411.4547",
    archivePrefix = "arXiv",
    primaryClass = "gr-qc",
    doi = "10.1088/0264-9381/32/7/074001",
    journal = "Class. Quant. Grav.",
    volume = "32",
    pages = "074001",
    year = "2015"
}

@article{VIRGO:2014yos,
    author = "Acernese, F. and others",
    collaboration = "VIRGO",
    title = "{Advanced Virgo: a second-generation interferometric gravitational wave detector}",
    eprint = "1408.3978",
    archivePrefix = "arXiv",
    primaryClass = "gr-qc",
    doi = "10.1088/0264-9381/32/2/024001",
    journal = "Class. Quant. Grav.",
    volume = "32",
    number = "2",
    pages = "024001",
    year = "2015"
}

@article{KAGRA:2020tym,
    author = "Akutsu, T. and others",
    collaboration = "KAGRA",
    title = "{Overview of KAGRA: Detector design and construction history}",
    eprint = "2005.05574",
    archivePrefix = "arXiv",
    primaryClass = "physics.ins-det",
    doi = "10.1093/ptep/ptaa125",
    journal = "PTEP",
    volume = "2021",
    number = "5",
    pages = "05A101",
    year = "2021"
}

@article{LIGOScientific:2018mvr,
    author = "Abbott, B. P. and others",
    collaboration = "LIGO Scientific, Virgo",
    title = "{GWTC-1: A Gravitational-Wave Transient Catalog of Compact Binary Mergers Observed by LIGO and Virgo during the First and Second Observing Runs}",
    eprint = "1811.12907",
    archivePrefix = "arXiv",
    primaryClass = "astro-ph.HE",
    reportNumber = "LIGO-P1800307",
    doi = "10.1103/PhysRevX.9.031040",
    journal = "Phys. Rev. X",
    volume = "9",
    number = "3",
    pages = "031040",
    year = "2019"
}

@article{LIGOScientific:2020aai,
    author = "Abbott, B. P. and others",
    collaboration = "LIGO Scientific, Virgo",
    title = "{GW190425: Observation of a Compact Binary Coalescence with Total Mass $\sim 3.4 M_{\odot}$}",
    eprint = "2001.01761",
    archivePrefix = "arXiv",
    primaryClass = "astro-ph.HE",
    reportNumber = "LIGO-P190425",
    doi = "10.3847/2041-8213/ab75f5",
    journal = "Astrophys. J. Lett.",
    volume = "892",
    number = "1",
    pages = "L3",
    year = "2020"
}

@article{Farr:2015lna,
    author = "Farr, Ben and others",
    title = "{Parameter estimation on gravitational waves from neutron-star binaries with spinning components}",
    eprint = "1508.05336",
    archivePrefix = "arXiv",
    primaryClass = "astro-ph.HE",
    reportNumber = "LIGO-P1500117",
    doi = "10.3847/0004-637X/825/2/116",
    journal = "Astrophys. J.",
    volume = "825",
    number = "2",
    pages = "116",
    year = "2016"
}

@article{Flanagan:2007ix,
    author = "Flanagan, Eanna E. and Hinderer, Tanja",
    title = "{Constraining neutron star tidal Love numbers with gravitational wave detectors}",
    eprint = "0709.1915",
    archivePrefix = "arXiv",
    primaryClass = "astro-ph",
    doi = "10.1103/PhysRevD.77.021502",
    journal = "Phys. Rev. D",
    volume = "77",
    pages = "021502",
    year = "2008"
}

@article{Hinderer:2007mb,
    author = "Hinderer, Tanja",
    title = "{Tidal Love numbers of neutron stars}",
    eprint = "0711.2420",
    archivePrefix = "arXiv",
    primaryClass = "astro-ph",
    doi = "10.1086/533487",
    journal = "Astrophys. J.",
    volume = "677",
    pages = "1216--1220",
    year = "2008",
    note = "[Erratum: Astrophys.J. 697, 964 (2009)]"
}

@article{Lindblom:2010bb,
    author = "Lindblom, Lee",
    title = "{Spectral Representations of Neutron-Star Equations of State}",
    eprint = "1009.0738",
    archivePrefix = "arXiv",
    primaryClass = "astro-ph.HE",
    doi = "10.1103/PhysRevD.82.103011",
    journal = "Phys. Rev. D",
    volume = "82",
    pages = "103011",
    year = "2010"
}

@article{Landry:2018prl,
    author = "Landry, Philippe and Essick, Reed",
    title = "{Nonparametric inference of the neutron star equation of state from gravitational wave observations}",
    eprint = "1811.12529",
    archivePrefix = "arXiv",
    primaryClass = "gr-qc",
    doi = "10.1103/PhysRevD.99.084049",
    journal = "Phys. Rev. D",
    volume = "99",
    number = "8",
    pages = "084049",
    year = "2019"
}

@article{Ng:2025wdj,
    author = "Ng, Sunny and Legred, Isaac and Suleiman, Lami and Landry, Philippe and Traylor, Lyla and Read, Jocelyn",
    title = "{Inferring the neutron star equation of state with nuclear-physics informed semiparametric models}",
    eprint = "2507.03232",
    archivePrefix = "arXiv",
    primaryClass = "astro-ph.HE",
    doi = "10.1088/1361-6382/ae1094",
    journal = "Class. Quant. Grav.",
    volume = "42",
    number = "20",
    pages = "205008",
    year = "2025"
}

@article{Magnall:2025zhm,
    author = "Magnall, Spencer J. and Ecker, Christian and Rezzolla, Luciano and Lasky, Paul D. and Goode, Simon R.",
    title = "{Physics-informed Priors Improve Gravitational-wave Constraints on Neutron-star Matter}",
    eprint = "2504.21526",
    archivePrefix = "arXiv",
    primaryClass = "astro-ph.HE",
    doi = "10.3847/2041-8213/adef39",
    journal = "Astrophys. J. Lett.",
    volume = "988",
    number = "2",
    pages = "L75",
    year = "2025"
}

@article{Pratten:2020fqn,
    author = "Pratten, Geraint and Husa, Sascha and Garcia-Quiros, Cecilio and Colleoni, Marta and Ramos-Buades, Antoni and Estelles, Hector and Jaume, Rafel",
    title = "{Setting the cornerstone for a family of models for gravitational waves from compact binaries: The dominant harmonic for nonprecessing quasicircular black holes}",
    eprint = "2001.11412",
    archivePrefix = "arXiv",
    primaryClass = "gr-qc",
    reportNumber = "LIGO-P2000018",
    doi = "10.1103/PhysRevD.102.064001",
    journal = "Phys. Rev. D",
    volume = "102",
    number = "6",
    pages = "064001",
    year = "2020"
}

@article{Abac:2023ujg,
    author = "Abac, Adrian and Dietrich, Tim and Buonanno, Alessandra and Steinhoff, Jan and Ujevic, Maximiliano",
    title = "{New and robust gravitational-waveform model for high-mass-ratio binary neutron star systems with dynamical tidal effects}",
    eprint = "2311.07456",
    archivePrefix = "arXiv",
    primaryClass = "gr-qc",
    doi = "10.1103/PhysRevD.109.024062",
    journal = "Phys. Rev. D",
    volume = "109",
    number = "2",
    pages = "024062",
    year = "2024"
}

@article{Blanchet:2013haa,
    author = "Blanchet, Luc",
    title = "{Post-Newtonian Theory for Gravitational Waves}",
    eprint = "1310.1528",
    archivePrefix = "arXiv",
    primaryClass = "gr-qc",
    doi = "10.12942/lrr-2014-2",
    journal = "Living Rev. Rel.",
    volume = "17",
    pages = "2",
    year = "2014"
}

@article{Fonseca:2021wxt,
    author = "Fonseca, E. and others",
    title = "{Refined Mass and Geometric Measurements of the High-mass PSR J0740+6620}",
    eprint = "2104.00880",
    archivePrefix = "arXiv",
    primaryClass = "astro-ph.HE",
    doi = "10.3847/2041-8213/ac03b8",
    journal = "Astrophys. J. Lett.",
    volume = "915",
    number = "1",
    pages = "L12",
    year = "2021"
}

@article{Vinciguerra:2023qxq,
    author = "Vinciguerra, Serena and others",
    title = "{An Updated Mass{\textendash}Radius Analysis of the 2017{\textendash}2018 NICER Data Set of PSR J0030+0451}",
    eprint = "2308.09469",
    archivePrefix = "arXiv",
    primaryClass = "astro-ph.HE",
    doi = "10.3847/1538-4357/acfb83",
    journal = "Astrophys. J.",
    volume = "961",
    number = "1",
    pages = "62",
    year = "2024"
}

@article{Choudhury:2024xbk,
    author = "Choudhury, Devarshi and others",
    title = "{A NICER View of the Nearest and Brightest Millisecond Pulsar: PSR J0437{\textendash}4715}",
    eprint = "2407.06789",
    archivePrefix = "arXiv",
    primaryClass = "astro-ph.HE",
    doi = "10.3847/2041-8213/ad5a6f",
    journal = "Astrophys. J. Lett.",
    volume = "971",
    number = "1",
    pages = "L20",
    year = "2024"
}

@article{Mauviard:2025dmd,
    author = "Mauviard, Lucien and others",
    title = "{A NICER View of the 1.4 M$_\odot$ Edge-on Pulsar PSR J0614-3329}",
    eprint = "2506.14883",
    archivePrefix = "arXiv",
    primaryClass = "astro-ph.HE",
    doi = "10.3847/1538-4357/ae145d",
    journal = "Astrophys. J.",
    volume = "995",
    number = "1",
    pages = "60",
    year = "2025"
}

@article{Salmi:2024bss,
    author = "Salmi, Tuomo and others",
    title = "{A NICER View of PSR J1231{\ensuremath{-}}1411: A Complex Case}",
    eprint = "2409.14923",
    archivePrefix = "arXiv",
    primaryClass = "astro-ph.HE",
    doi = "10.3847/1538-4357/ad81d2",
    journal = "Astrophys. J.",
    volume = "976",
    number = "1",
    pages = "58",
    year = "2024"
}

@article{Dittmann:2024mbo,
    author = "Dittmann, Alexander J. and others",
    title = "{A More Precise Measurement of the Radius of PSR J0740+6620 Using Updated NICER Data}",
    eprint = "2406.14467",
    archivePrefix = "arXiv",
    primaryClass = "astro-ph.HE",
    doi = "10.3847/1538-4357/ad5f1e",
    journal = "Astrophys. J.",
    volume = "974",
    number = "2",
    pages = "295",
    year = "2024"
}

@article{Romani:2025ytn,
    author = "Romani, Roger W. and Beleznay, Maya and Filippenko, Alexei V. and Brink, Thomas G. and Zheng, WeiKang",
    title = "{PSR J0952-0607: Tightening a Record-high Neutron Star Mass}",
    eprint = "2512.05099",
    archivePrefix = "arXiv",
    primaryClass = "astro-ph.HE",
    doi = "10.3847/1538-4357/ae28c5",
    journal = "Astrophys. J.",
    volume = "996",
    number = "1",
    pages = "101",
    year = "2026"
}

@article{Romani:2021xmb,
    author = "Romani, Roger W. and Kandel, D. and Filippenko, Alexei V. and Brink, Thomas G. and Zheng, WeiKang",
    title = "{PSR J1810+1744: Companion Darkening and a Precise High Neutron Star Mass}",
    eprint = "2101.09822",
    archivePrefix = "arXiv",
    primaryClass = "astro-ph.HE",
    doi = "10.3847/2041-8213/abe2b4",
    journal = "Astrophys. J. Lett.",
    volume = "908",
    number = "2",
    pages = "L46",
    year = "2021"
}

@article{Saffer:2024tlb,
    author = "Saffer, Alexander and others",
    title = "{A Lower Mass Estimate for PSR J0348+0432 Based on CHIME/Pulsar Precision Timing}",
    eprint = "2412.02850",
    archivePrefix = "arXiv",
    primaryClass = "astro-ph.HE",
    doi = "10.3847/2041-8213/adc25e",
    journal = "Astrophys. J. Lett.",
    volume = "983",
    number = "1",
    pages = "L20",
    year = "2025"
}

@article{Wouters:2025zju,
    author = "Wouters, Thibeau and Pang, Peter T. H. and Koehn, Hauke and Rose, Henrik and Somasundaram, Rahul and Tews, Ingo and Dietrich, Tim and Van Den Broeck, Chris",
    title = "{Leveraging differentiable programming in the inverse problem of neutron stars}",
    eprint = "2504.15893",
    archivePrefix = "arXiv",
    primaryClass = "astro-ph.HE",
    reportNumber = "LA-UR-25-23486",
    doi = "10.1103/v2y8-kxvx",
    journal = "Phys. Rev. D",
    volume = "112",
    number = "4",
    pages = "043037",
    year = "2025"
}

@article{Baade:1934zex,
  author = {Baade, W. and Zwicky, F.},
  title = {{On Super-Novae}},
  doi = {10.1073/pnas.20.5.254},
  journal = {Proc. Nat. Acad. Sci.},
  volume = {20},
  number = {5},
  pages = {254--259},
  year = {1934},
}

@article{Chatziioannou:2020pqz,
    author = "Chatziioannou, Katerina",
    title = "{Neutron star tidal deformability and equation of state constraints}",
    eprint = "2006.03168",
    archivePrefix = "arXiv",
    primaryClass = "gr-qc",
    doi = "10.1007/s10714-020-02754-3",
    journal = "Gen. Rel. Grav.",
    volume = "52",
    number = "11",
    pages = "109",
    year = "2020"
}

@article{Favata:2013rwa,
    author = "Favata, Marc",
    title = "{Systematic parameter errors in inspiraling neutron star binaries}",
    eprint = "1310.8288",
    archivePrefix = "arXiv",
    primaryClass = "gr-qc",
    reportNumber = "LIGO-P1300191",
    doi = "10.1103/PhysRevLett.112.101101",
    journal = "Phys. Rev. Lett.",
    volume = "112",
    pages = "101101",
    year = "2014"
}

@article{Wouters:2025csq,
  author = {Wouters, Thibeau and Pang, Peter T. H. and Dietrich, Tim and Van Den Broeck, Chris},
  title = {{Incorporating neutron star physics into gravitational wave inference with physics-informed priors using normalizing flows}},
  eprint = {2511.22987},
  archiveprefix = {arXiv},
  primaryclass = {astro-ph.HE},
  doi = {10.1103/8t5h-y4lf},
  journal = {Phys. Rev. D},
  volume = {114},
  number = {4},
  pages = {043005},
  year = {2026},
}

@article{Pratten:2020ceb,
  author = {Pratten, Geraint and others},
  title = {{Computationally efficient models for the dominant and subdominant harmonic modes of precessing binary black holes}},
  eprint = {2004.06503},
  archiveprefix = {arXiv},
  primaryclass = {gr-qc},
  doi = {10.1103/PhysRevD.103.104056},
  journal = {Phys. Rev. D},
  volume = {103},
  number = {10},
  pages = {104056},
  year = {2021},
}

@unpublished{Prasad:2026idz,
  author = {Prasad, Vaishak and Sathyaprakash, B. S.},
  title = {{Are black hole spins truly near-zero?}},
  eprint = {2602.12859},
  archiveprefix = {arXiv},
  primaryclass = {gr-qc},
  reportnumber = {LIGO-P2500751},
  month = {2},
  year = {2026},
}

@article{Finn:1992wt,
  author = {Finn, Lee S.},
  title = {{Detection, measurement and gravitational radiation}},
  eprint = {gr-qc/9209010},
  archiveprefix = {arXiv},
  reportnumber = {PRINT-93-0128 (NORTHWESTERN)},
  doi = {10.1103/PhysRevD.46.5236},
  journal = {Phys. Rev. D},
  volume = {46},
  pages = {5236--5249},
  year = {1992},
}
